\documentclass{svproc}
\usepackage{url}

\usepackage{hyperref}
\usepackage{type1cm}
\usepackage{makeidx}
\usepackage{graphicx}
\graphicspath{{figures/}}

\usepackage{multicol}
\usepackage[bottom]{footmisc}

\usepackage{newtxtext}
\usepackage[varvw]{newtxmath}

\usepackage{bm}
\usepackage{siunitx}
\usepackage{anyfontsize}
\usepackage{mathtools}
\usepackage{subcaption}
\usepackage{float}
\usepackage{placeins}

\newcounter{algorithm}

\newcommand{\bsb}{{\boldsymbol{b}}}

\newcommand{\bsg}{{\boldsymbol{g}}}

\newcommand{\bsm}{{\boldsymbol{m}}}
\newcommand{\bsn}{{\boldsymbol{n}}}

\newcommand{\bsp}{{\boldsymbol{p}}}

\newcommand{\bsr}{{\boldsymbol{r}}}

\newcommand{\bsu}{{\boldsymbol{u}}}
\newcommand{\bsv}{{\boldsymbol{v}}}

\newcommand{\bsx}{{\boldsymbol{x}}}
\newcommand{\bsy}{{\boldsymbol{y}}}

\newcommand{\bsB}{{\boldsymbol{B}}}
\newcommand{\bsC}{{\boldsymbol{C}}}
\newcommand{\bsD}{{\boldsymbol{D}}}

\newcommand{\bsF}{{\boldsymbol{F}}}

\newcommand{\bsH}{{\boldsymbol{H}}}

\newcommand{\bsM}{{\boldsymbol{M}}}

\newcommand{\bsV}{{\boldsymbol{V}}}
\newcommand{\bsW}{{\boldsymbol{W}}}
\newcommand{\bsX}{{\boldsymbol{X}}}

\newcommand{\bseta}{{\boldsymbol{\eta}}}

\newcommand{\bsxi}{{\boldsymbol{\xi}}}

\newcommand{\bsGamma}{{\boldsymbol{\Gamma}}}

\newcommand{\bsSigma}{{\boldsymbol{\Sigma}}}

\newcommand{\bbE}{{\mathbb{E}}}

\newcommand{\bbN}{{\mathbb{N}}}

\newcommand{\bbR}{{\mathbb{R}}}

\DeclareSymbolFont{bbold}{U}{bbold}{m}{n}
\DeclareSymbolFontAlphabet{\mathbbold}{bbold}

\newcommand{\calA}{{\mathcal{A}}}
\newcommand{\calB}{{\mathcal{B}}}
\newcommand{\calC}{{\mathcal{C}}}

\newcommand{\calF}{{\mathcal{F}}}

\newcommand{\calH}{{\mathcal{H}}}
\newcommand{\calI}{{\mathcal{I}}}

\newcommand{\calL}{{\mathcal{L}}}
\newcommand{\calM}{{\mathcal{M}}}
\newcommand{\calN}{{\mathcal{N}}}
\newcommand{\calO}{{\mathcal{O}}}
\newcommand{\calP}{{\mathcal{P}}}

\newcommand{\calR}{{\mathcal{R}}}

\newcommand{\calU}{{\mathcal{U}}}

\providecommand{\argmin}{\operatorname*{argmin}}

\begin{document}
\mainmatter

\title{Non-parametric Inference for Diffusion Processes: \\ A Computational Approach via Bayesian Inversion \\ for PDEs}
\titlerunning{Diffusion Process Inference}

\author{Maximilian Kruse\inst{1} \and Sebastian Krumscheid\inst{1}}
\authorrunning{Kruse and Krumscheid}
\tocauthor{Maximilian Kruse, Sebastian Krumscheid}

\institute{Karlsruhe Institute of Technology,\\
\email{maximilian.kruse@kit.edu},\\
\email{sebastian.krumscheid@kit.edu}}

\maketitle

\begin{abstract}
In this paper, we present a theoretical and computational workflow for the non-parametric Bayesian inference of drift and diffusion functions of autonomous diffusion processes. We base the inference on the partial differential equations arising from the infinitesimal generator of the underlying process. Following a problem formulation in the infinite-dimensional setting, we discuss optimization- and sampling-based solution methods. As preliminary results, we showcase the inference of a single-scale, as well as a multiscale process from trajectory data.
\end{abstract}


\section{Introduction}\label{sec:introduction}

Various phenomena in science and engineering can be modelled via stochastic processes, or more precisely, diffusion processes. Applications include, to mention a few, problems in biology, climate science, energy technology, and finance (see, e.g., \cite{pavliotis_stochastic_2014} and the references therein). More recently, diffusion processes have found their use in machine learning, such as in deep generative models \cite{yang_diffusion_2024}. A special use-case is the description of interacting many-particle systems, as they occur, for instance, in molecular dynamics simulations. Such systems often allow for an effective coarse-grain representation via stochastic dynamics. 

Diffusion processes need to be parameterized by suitable drift and diffusion functions, which are typically unknown from first principles. On the other hand, a wealth of data is often available in the form of stochastic trajectories. Indeed, computational procedures such as molecular dynamics simulations typically generate entire ensembles of trajectories \cite{knapp_avoiding_2018}. Calibration of the parameter functions from this indirect data is therefore an important endeavor.

As drift and diffusion have to be considered as functions defined on a system's state space, they are formally infinite-dimensional objects. Thus, their inference from trajectory data, which is mostly available at discrete points in space and time, is severely ill-posed \cite{papaspiliopoulos_nonparametric_2012,croix_nonparametric_2020}. A possible remedy is parametric inference \cite{krumscheid_semiparametric_2013,krumscheid_data-driven_2015,craigmile_statistical_2023}, but this imposes possibly strong a-priori assumptions on the unknown quantities. In addition, the available data is commonly noise-polluted which, in combination with modelling assumptions, introduces uncertainty into the parameter estimates. A Bayesian framework provides an elegant approach to addressing both of these issues. By treating both the data and unknowns as random variables, we can naturally incorporate uncertainties into the inference process. The definition of a prior measure further provides a function space regularizer that incorporates a-priori knowledge about the parameter functions, i.e., the drift and diffusion.

In this work, we present a workflow for non-parametric Bayesian inference for the parameter functions of diffusion processes. We base our exposition on the function space viewpoint presented in \cite{stuart_inverse_2010}. For the forward model, we exploit the link of a diffusion process with its generator via the Kolmogorov partial differential equations, which govern the evolution of the process's observables and probability densities, respectively.
We further utilize a collection of methods for the assessment of the posterior measure that is specifically tailored towards large-scale problems \cite{ghattas_learning_2021,villa_hippylib_2021}. The resulting computational pipeline is demonstrated through inference for a single-scale process from trajectory data. Lastly, we infer an effective coarse-grain representation of a multiscale process.
\section{Methods}\label{sec:methods}

\subsection{Stochastic Processes and PDE Representation}\label{sec:sde_theory}

In the following, we are concerned with the inference of parameter functions for time-homogeneous diffusion processes on $\Omega\subseteq\bbR^d$, continuously indexed over $t\in\bbR_+$. Such processes are typically written as an Itô stochastic differential equation (SDE),
\begin{equation}
    d\bsX_t = \bsb(\bsX_t)dt + \sqrt{\bsSigma(\bsX_t)} d\bsW_t,\quad\bsX(t=0)=\bsX_0\ \ a.s.,
    \label{eq:ito_sde}
\end{equation}
where $\bsX_t \in\Omega$ is the process arising from the solution of the SDE, $\bsb: \Omega \to \bbR^d$ and $\bsSigma: \Omega \to \bbR^{d\times d}$ are the state-dependent drift vector and symmetric positive definite (s.p.d) diffusion tensor, respectively, and $\bsW_t\in\bbR^d$ is a Wiener process. For simplicity, we have restricted the formulation to the case of a square diffusion tensor, but this is not a strict requirement for what follows

Diffusion processes further admit a representation in terms of the Kolmogorov equations, a pair of parabolic partial differential equations (PDE) \cite{pavliotis_stochastic_2014,gardiner_handbook_2004}. Specifically, we define the generator of a process as a second order differential operator,
\begin{equation}
    \calL = \bsb(\bsx)\cdot \nabla + \frac{1}{2}\bsSigma(\bsx)\colon\nabla\nabla.
\end{equation}
The law $p: \Omega\times\bbR_+ \to \bbR_+$, with $p \geq 0, ||p(\cdot,t)||_{L^1} = 1\ \forall t$, of the process is then governed by the Kolmogorov forward or Fokker-Planck equation,
\begin{equation}
    \frac{\partial p}{\partial t}(\bsx,t) = \calL^* p(\bsx,t), \quad p(\bsx,t=0) = p_0(\bsx),
\end{equation}
where $\calL^*$ denotes the formal $L^2$-adjoint of the generator $\calL$. We assume that $p$ is a valid probability density function (PDF) with respect to the Lebesgue measure. On the other hand, the evolution of any continuous bounded observable on the state space, $f\in C_b(\Omega)$, is characterized in expectation by the Kolmogorov backward equation,
\begin{subequations}
\begin{gather}
    u(\bsx, t)\coloneqq \bbE[f(\bsX_t)|\bsX_0=\bsx], \\
    \frac{\partial u}{\partial t}(\bsx,t) = \calL u(\bsx,t), \quad u(\bsx,t=0) = f(\bsx),
\end{gather}
\end{subequations}
equipped with suitable boundary conditions.
In particular, the Kolmogorov backward equation gives rise to a stationary PDE whose solution is the Mean Exit Time or Mean First Passage Time (MFPT) with respect to a bounded subset $\calA$ of the domain $\Omega$. The first passage time (FPT) of a trajectory through the respective boundary $\partial\calA$ is defined as
\begin{equation}
    \tau(x) \coloneqq \inf\{ t\geq 0: \bsX_t \notin \calA | \bsX_0 = \bsx \}
\end{equation}
The moments of the FTP distribution, $\tau_n(\bsx)\coloneqq\bbE[\tau^n(\bsx)]$, can be obtained from a recursive hierarchy of PDEs. For absorbing boundaries $\partial\calA$, the MFPT $\tau_1$ is given as
\begin{subequations}
\begin{gather}
    \calL\tau_1 = -1,\quad x\in\calA, \\
    \tau_1 = 0,\quad x\in\partial\calA.
\end{gather}
\end{subequations}
Higher moments are evaluated as
\begin{subequations}
\begin{gather}
    \calL\tau_n = -n\tau_{n-1},\quad x\in\calA, \\
    \tau_n = 0,\quad x\in\partial\calA.
\end{gather}
\end{subequations}

\subsection{Bayesian Inverse Problem Formulation}\label{sec:bip_setup}

Given the previously described PDE models, we aim to solve the inverse problem of finding the unknown drift and diffusion functions of a given process from PDF or MFPT data. We follow the Bayesian framework put forward in \cite{stuart_inverse_2010,sullivan_introduction_2015,ghanem_handbook_2017}, which adopts a function space viewpoint and defers discretization to the latest possible moment. This allows for discretization-invariant analysis and solution methods, as discussed later.

Formally, we define the unknown parameter as $\bsm = \text{vec}(\bsb, \bsSigma) \in \calM$ with a total of $M\in\bbN$ scalar components. In the course of this work, we assume for simplicity that $\bsm$ is a member of a product Hilbert space, $\calM = \bigotimes_{i=1}^M \big(\calH^1(\Omega)\big)_i$, composed of the scalar components of $\bsb$ and $\bsSigma$. We note, however, that our methods are not restricted to this scenario. We further assume data for MFPT moments to be available at discrete locations $\bsx_1, \bsx_2,\ldots, \bsx_q$ in state space, meaning we have for the n-th moment $\bsy_n = (y_{n,1}, y_{n,2},\ldots,y_{n,q})\in\bbR^q$. We collect all data up to the k-th moment in a global vector $\bsy = \text{vec}(\bsy_1,\bsy_2,\ldots,\bsy_k)\in\bbR^{kq}$. Regarding PDF data, we assume for simplicity the same discrete locations in space, for a finite number of time points $t_1,t_2,\ldots,t_l$. This results in a global data vector $\bsy\in\bbR^{lq}$.

Unknowns and data are related via a parameter-to-observable (PTO) mapping, which consists of two sequential steps. Firstly, the parameters are related to solution variables $\bsu\in\calU$ via the corresponding PDE constraint, which we abstractly denote as $r:\calU \times \calM \to \calU^*,\ r(\bsu,\bsm) = 0$, with $\calU^*$ being the dual space of $\calU$. In the case of the MFPT equations for moments up to order $k$, we have $\bsu = \text{vec}(\tau_1,\tau_2,\ldots,\tau_k)\in\calU_\text{MFPT} = \bigotimes_{n=1}^k\big(\calH_0^1(\Omega)\big)_n$. For the Fokker-Planck equation, we pose the solution in the Bochner space $\calU_\text{FP} = \calL^2\big(\calR_+;\calH^1(\Omega)\big)$.
Subsequently, the solution can be projected onto the points (and times) of observation through a projection operator $\calB$. Specifically, we write $\calB_\text{MFPT}: \calU_\text{MFPT}\to\bbR^{kq}$ and $\calB_\text{FP}: \calU_\text{FP}\to\bbR^{lq}$. In total, we define a generic PTO mapping as
\begin{equation}
    \calF(\bsm) = \calB(\bsu),\quad \text{s.t.}\ \ r(\bsu,\bsm)=0.
\end{equation}

In the Bayesian setting, we treat both unknowns and data as random variables. With regard to the former, this naturally accounts for the fact that the observations are typically noise-polluted instances of the true observables. For the current work, we simply postulate additive, zero-centered Gaussian noise on the data and write
\begin{equation}
    \bsy = \calF(\bsm) + \bseta, \quad \bseta \sim \calN(0, \bsGamma),
\end{equation}
with s.p.d noise covariance matrices $\bsGamma_\text{MFPT}\in\bbR^{(kq)\times(kq)}$ and $\bsGamma_\text{FP}\in\bbR^{(lq)\times(lq)}$, respectively. Importantly, our methods are easily extensible towards other noise models. For the present case, it follows that we can define a likelihood functional as
\begin{equation}
    \pi_\text{like}(\bsy | \bsm) \propto \exp\Big\{-\frac{1}{2}||\bsy-\calF(\bsm)||_\bsGamma^2\Big\},
\end{equation}
where we have introduced the matrix weighted norm $||\bsx||_\Gamma \coloneqq \sqrt{\bsx^T\Gamma^{-1}\bsx}$.

To complete the formulation of the Bayesian inverse problem (BIP), we define a prior measure as
\begin{equation}
    d\mu_\text{prior}(\bsm) \coloneqq \mu_\text{prior}(d\bsm) = \bigotimes_{i=1}^M\mu_{\text{prior},i}(dm_i).
\end{equation}
Independently for each scalar component $i=1,\ldots,M$, we assign a Gaussian measure $\calN(\overline{m}_i,\calC_i)$ with mean function $\overline{m}_i$ and trace-class, positive, self-adjoint covariance operator $\calC_i$. Such Gaussian fields are the most established class of priors in the infinite-dimensional setting. They allow for reasonable flexibility in spatial variance and correlation structures, as well as efficient computations.
For the covariance, we employ an elliptic differential operator, reminiscent of the inverse bi-Laplacian \cite{lindgren_explicit_2011,lindgren_spde_2022},
\begin{equation}
    \calC_i = \big(\delta_i(\bsx)\calI -\gamma_i(\bsx)\Delta\big)^{-2},\quad \bsx\in\Omega.
    \label{eq:bilaplacian_prior}
\end{equation}
We explicitly include the possibility of spatially varying parameters, which translates to non-stationary fields. These parameters have direct correspondence to the pointwise variance $\sigma^2$ and correlation length $\rho$,
\begin{equation}
    \sigma_i^2 = \frac{\Gamma(\nu)}{(4\pi)^{d/2}}\frac{1}{\delta_i^\nu\gamma_i^{d/2}},\quad \rho = \sqrt{\frac{8\nu\gamma_i}{\delta_i}},\quad \nu = 2-\frac{d}{2}.
\end{equation}
We further point out that $\nu$ indicates the order of mean-square differentiability of the resulting field, and that $\nu>0$ is required for the field to be well-defined. For practical computations, the PDE \eqref{eq:bilaplacian_prior} needs to be equipped with boundary conditions. We choose Robin boundary conditions to (partially) alleviate boundary artifacts in the PDE solution \cite{daon_mitigating_2018},
\begin{equation}
    \gamma_i\nabla m_i\cdot\bsn + \frac{\sqrt{\gamma_i\delta_i}}{1.42}m_i = 0,\quad \bsx\in\partial\Omega.
\end{equation}
The above choice for the prior field has two major advantages. Firstly, it obeys a continuous analog to the Markov property in discretely index fields. This leads to sparse precision matrices after discretization, enabling efficient computations. Secondly, the differential operator approach has a direct correspondence to Matérn type fields, whose statistical properties are well understood.

We can now state a generalized version of Bayes' theorem in the form of a Radon--Nikodym derivative,
\begin{equation}
    \frac{d\mu_\text{post}(\bsm|\bsy)}{d\mu_\text{prior}(\bsm)} \propto \pi_\text{like}(\bsy|\bsm),
\end{equation}
which defines (proportionality to) a posterior measure $d\mu_\text{post}$ for the unknown parameter functions $\bsm$ given the data $\bsy$.

\subsection{Solution Methods}\label{sec:solution_methods}
Given a formulation of the posterior measure, we intend to extract sensible information about the parameters' distribution from it. One possible approach is by means of variational methods. To this end, we firstly aim to determine the maximum a-posteriori (MAP) estimate. Again following the work in \cite{stuart_inverse_2010,ghanem_handbook_2017}, this is the infinite-dimensional analogue of determining the parameter candidate that maximizes the posterior density with respect to the Lebesgue measure. Let $\bsB(\epsilon,\bsm)\subset\calM$ be the open ball with radius $\epsilon > 0$ and centered at $\bsm\in\calM$. We define the MAP estimate as the parameter candidate that maximizes the posterior mass of that ball $\mu_\text{post}(\bsB(\epsilon,\bsm))$ as $\epsilon\to 0$. In the given setting, this yields a constrained optimization problem as
\begin{subequations}
\begin{gather}
    \bsm_\text{MAP} = \argmin_{\bsm\in\calM} J(\bsu,\bsm),\quad s.t.\ r(\bsu,\bsm) = 0, \\
    \text{with}\quad J(\bsu,\bsm) = \frac{1}{2}|| \calB(\bsu)-\bsy||_\bsGamma^2 + \frac{1}{2}||\bsm-\overline{\bsm}||_\calC^2.
\end{gather}
\label{eq:constrained_optimization_problem}
\end{subequations}
Note that, strictly speaking, the minimizer has to be found in the Cameron-Martin space associated with the prior measure. However, we disregard this technicality in view of a practical solution.
Relation \eqref{eq:constrained_optimization_problem} can be transformed into an unconstrained optimization problem by means of the formal Lagrangian method \cite{borzi_computational_2011,gunzburger_perspectives_2003,hinze_optimization_2009,troltzsch_optimal_2010}. We define the Lagrangian as a functional of three independent variables,
\begin{equation}
    \calL(\bsu,\bsm,\bsp) \coloneqq J(\bsu,\bsm) + \big(\bsp, r(\bsu,\bsm)\big)_{L^2(\Omega)},
\end{equation}
where we have introduced the Lagrange multiplier, or adjoint variable, $\bsp\in\calP=\calU^{**}$. We have further foregone the introduction of a more generic dual pairing, and directly work with the $L^2$ inner product by virtue of the Riesz representation theorem. Assuming sufficient (Fréchet) differentiability, we can state the first order necessary condition on the optimizer $\bsm_\text{MAP}$ that all partial first variations of the Lagrangian have to vanish for arbitrary test directions,
\begin{subequations}
\begin{align}
    \delta\calL_u(\bsu,\bsm,\bsp)[\Tilde{\bsu}] &= 0,\quad \forall\Tilde{\bsu}\in\calU \label{eq:adjoint_equation}\\
    \delta\calL_m(\bsu,\bsm,\bsp)[\Tilde{\bsm}] &= 0,\quad \forall\Tilde{\bsm}\in\calM \label{eq:forward_equation}\\
    \delta\calL_m(\bsu,\bsm,\bsp)[\Tilde{\bsp}] &= 0,\quad \forall\Tilde{\bsp}\in\calP \label{eq:control_equation}.
\end{align}   
\end{subequations}
These relations are referred to as adjoint, forward and control equations, respectively. For Newton-type optimization methods, we can formally define an update step as \cite{petra_model_2011}
\begin{equation}
    \delta^2\calL(\bsu,\bsm,\bsp)[(\Tilde{\bsu},\Tilde{\bsm},\Tilde{\bsp});(\hat{\bsu},\hat{\bsm},\hat{\bsp})] = - \delta\calL(\bsu,\bsm,\bsp)[(\Tilde{\bsu},\Tilde{\bsm},\Tilde{\bsp})],
    \label{eq:lagrange_second_variation}
\end{equation}
which yields update directions $\hat{\bsu}\in\calU$, $\hat{\bsm}\in\calM$, and $\hat{\bsp}\in\calP$.

We can further enrich the MAP estimate by means of the Laplace approximation, a local Gaussian approximation about that point \cite{bui-thanh_extreme-scale_2012,bui-thanh_computational_2013,ghattas_learning_2021}. The Laplace approximation is based on a linearization of the PTO map,
\begin{equation}
    \calF(\bsm) \approx \calF'(\bsm_\text{MAP})(\bsm-\bsm_\text{MAP}),
\end{equation}
where $\calF'$ denotes the Fréchet derivative of $\calF$. This approximation yields a Gaussian posterior with mean $\bsm_\text{MAP}$ and covariance operator
\begin{equation}
    \calC_\text{LA} = \Big[ \big(\calF'(\bsm_\text{MAP})\big)^*\bsGamma^{-1}\calF'(\bsm_\text{MAP}) + \calC^{-1} \Big]^{-1}.
    \label{eq:laplace_covariance}
\end{equation}
The covariance can also be seen as the inverse Hessian operator $\calH^{-1}(\bsm_\text{MAP})$ of the log-posterior at the MAP point.

The described variational method provides an approximation to the posterior measure that can give valuable insides on the solution of the BIP. However, it can hardly represent e.g. multimodal-distributions, skewness, or other forms of non-Gaussianity. To draw samples from the exact posterior, we therefore additionally resort to Markov Chain Monte Carlo (MCMC) methods. More specifically, we employ an algorithm from the class of dimension-independent Metropolis-Hastings (MH) methods \cite{tierney_note_1998,cotter_mcmc_2013,beskos_geometric_2017}. These methods are constructed from proposals that are reversible with respect to a (Gaussian) reference measure, which does not necessarily have a Lebesgue density. In contrast to conventional MCMC samplers, they do not suffer from degeneration of mixing times under mesh refinement. In our work, we employ an infinite-dimensional generalization of the Metropolis-Adjusted Langevin (MALA) sampler. This sampler further incorporates information on the geometry of the posterior measure in the form of the covariance $\calC_\text{LA}$. The overall proposal for the MH-Sampler then reads
\begin{equation}
    \Tilde{\bsm} = \rho\bsm + \sqrt{\frac{h(1-\rho^2)}{4}}\Big(\bsm-\calC_\text{LA}\calC^{-1}(\bsm-\overline{\bsm})-\calC_\text{LA}\Phi'(\bsm)\Big) + \sqrt{1-\rho^2}\bsxi,
    \label{eq:mala_proposal}
\end{equation}
with step width parameter $h>0$, $\rho = \frac{4-h}{4+h}$, and $\bsxi\sim\calN(0,\calC_\text{LA})$. We have further introduced the negative log-likelihood functional $\Phi(\bsm) = \frac{1}{2}|| \calF(\bsm) -\bsy||_\bsGamma^2$ depending directly on $\bsm$ and for fixed data $\bsy$.
\section{Computational Aspects}\label{sec:numerics}

The computational realization of the described solution methods clearly requires discretization of the underlying function space objects. This yields a high- but finite-dimensional BIP. In the following, we discuss a class of algorithms that, similar to the presented MCMC method, is designed with the infinite-dimensional nature of the continuous problem in mind \cite{bui-thanh_extreme-scale_2012,bui-thanh_computational_2013,petra_model_2011,ghattas_learning_2021,nocedal_numerical_2006}. In particular, these procedures exploit the spectral properties of the underlying operators. This yields computational methods with favorable complexity, concerning operations as well as memory usage, for a large number of degrees of freedom.

\subsection{Discretization}
A consistent discretization of the involved equations is a prerequisite for the employed computational methods to work properly. To this end, we employ the finite element method (FEM) for discretization in space, and the finite difference method (FDM) for time discretization.
Consistency is primarily a concern with respect to space, so we restrict our exposition to a stationary quantity $\bsm\in\calM$. Following standard FEM procedure \cite{zienkiewicz_finite_2010}, we approximate $\bsm$ on a finite-dimensional subspace $\calM_h\subset\calM$ as a linear combination of basis functions $\{\phi_i\}_{i=1}^N$ of $\calM_h$,
\begin{equation}
    \hat{m}(\bsx) = \sum_{i=1}^N m_i\phi_i(\bsx) \in\calM_h.
\end{equation}
Regarding the coefficient vector $\bsm_h \coloneqq (m_1, m_2, \ldots, m_N) \in\bbR^N$, we can define a discrete approximation of the $L^2$ inner product. Let $\bsm,\bsn\in\calM$, then it follows that
\begin{equation}
    (\bsm,\bsn)_{L^2(\Omega)} \approx (\hat{\bsm},\hat{\bsn})_{L^2(\Omega)} = (\bsm_h,\bsM\bsn_h)_{\bbR^N} \eqqcolon (\bsm_h,\bsn_h)_\bsM,
\end{equation}
where we have introduced the mass matrix $\bsM\in\bbR^{N\times N}$ with $M_{ij} = \int_\Omega\phi_i(x)\phi_j(x)dx$. Discretization of variational forms, operators, and their adjoints is conducted with respect to the mass matrix weighted inner product. This ensures mesh-independence of operator properties, such as correlation lengths of the prior and posterior fields.

\subsection{Maximum a-posteriori Estimate}
Given a discretization of the BIP, we turn to the numerical evaluation of the MAP point. Again assuming that these quantities exist, we define the total gradient of the discrete MAP cost functional \eqref{eq:constrained_optimization_problem} as $\bsg(\bsm_h)\in\bbR^N$, and its Hessian as $\bsH(\bsm_h)\in\bbR^{N\times N}$. Equation \eqref{eq:lagrange_second_variation} can then be condensed into a discrete Newton update \cite{petra_model_2011}
\begin{equation}
    \bsH(\bsm_{h,k})\hat{\bsm}_{h,k} = -\bsg(\bsm_{h,k})
\end{equation}
for an update direction $\hat{\bsm}_{h,k}\in\bbR^N$, and some iteration $k$. These Newton iterations can be performed efficiently under the assumption that the (discretization of) the Hessian operator is compact. For many BIPs, this assumption is justified, for instance in the case of sparse data, or considering that highly oscillatory components in the PTO map can often be neglected. Furthermore, compactness can be enhanced through preconditioning with the prior precision operator. Hence, we work with the assumption that $\bsH$ has a ``dominant'' rank $r$, beyond which all its eigenvalues are small, $\lambda_j \ll 1,\ \forall j > r$, with $r\ll N$.

Under the compactness assumption, the inexact Newton-CG method provides a scalable optimization algorithm \cite{ghattas_learning_2021}. Linear solves are conducted matrix-free using the conjugate gradient (CG) method. According to the Lagrangian formalism, Hessian-vector products can be computed with computational cost proportional to that of a forward PDE solve, while never assembling the (dense) Hessian matrix. Furthermore, the CG method firstly penalizes error components associated with the subspace of dominant eigenvectors of the system matrix. Thus, the ``dominant" contributions to the residual $\bsr_k = \bsH(\bsm_{h,k})\hat{\bsm}_{h,k} + \bsg(\bsm_{h,k}) $ are eliminated in $\calO(r)$ CG iterations. The inexact Newton-CG method further employs early stopping of the CG iterations, with the Eisenstat--Walker termination criterion $||\bsr_k|| \leq \eta_k ||\bsg(\bsm_{h,k})||$ \cite{villa_hippylib_2021}. The prefactor $\eta_k$ itself depends on $\bsg$ and determines the convergence rate of the Newton iterations. Early stopping ensures a number of CG iterations independent of $N$, and further avoids over-solving far away from the optimum. To ensure a descent direction, the employed algorithm further makes use of the Steihaug criterion, and facilitates global convergence through Armijo line search. Lastly, we note that the outer Newton iterations are known to converge to a given tolerance independently of the problem size for a wide range of applications, making the optimization algorithm overall scalable.

\subsection{Laplace Approximation}
Compactness of the Hessian also allows for the efficient construction of the covariance for the Laplace approximation. According to equation \eqref{eq:laplace_covariance}, we can decompose the expression for the MAP Hessian as $\bsH_\text{MAP} = \bsH_\text{MAP}^\text{Data} + \bsC^{-1}$ with $\bsH_\text{MAP}^\text{Data}=(\bsF_\text{MAP}')^T\bsGamma^{-1}\bsF_\text{MAP}'$. Assuming that $\bsH_\text{MAP}^\text{Data}$ is compact with dominant rank $r\ll N$, we can conduct a truncated generalized eigenvalue decomposition,
\begin{equation}
    \bsH_\text{MAP}^\text{Data}\bsv_j = \lambda_j\bsC^{-1}\bsv_j,\quad j=1,2,\ldots,r.
\end{equation}
Note that $\bsC^{-1}$ is readily available as the (sparse) discretization of the precision operator. The eigenvalue decomposition can be performed matrix-free and with $\calO(r)$ Hessian-vector products using randomized linear algebra methods \cite{halko_finding_2011,martinsson_randomized_2020,murray_randomized_2023}. Now defining the matrices
\begin{equation}
    \bsV_r = (\bsv_1,\bsv_2,\ldots,\bsv_r)\in\bbR^{N\times r}, \quad
    \bsD = \text{diag}\bigg(\frac{\lambda_j}{\lambda_j+1}\bigg)\in\bbR^{r\times r},\ j=1,2,\ldots,r
\end{equation}
we can formulate the Laplace covariance through the Sherman-Morrison-Woodbury identity,
\begin{equation}
    \bsC_\text{LA} = \bsC - \bsV_r\bsD_r\bsV_r^T + \calO\bigg(\sum_{j=r+1}^N\frac{\lambda_j}{\lambda_j+1}\bigg),
\end{equation}
where the trailing eigenvalue term is small and can be omitted. We have therefore arrived at a scalable low-rank approximation, which can also be employed for the MCMC proposal \eqref{eq:mala_proposal}.

\subsection{Implementation}
For most of the described computational procedures, we have utilized the \texttt{hIPPYlib} software library\cite{villa_hippylib_2021}, a collection of tools for the solution of large-scale (Bayesian) inverse problems. It provides algorithms for computations involving prior fields, likelihoods, Lagrange systems, optimization, and low-rank constructions of the Laplace approximation. We have extended \texttt{hIPPYlib} for our work with capabilities for time-dependent PDE solves, joint inversion of vector and tensor fields, and non-stationary prior fields. For MCMC sampling, we have relied on the \texttt{MUQ} library \cite{parno_muq_2021} via the \texttt{hIPPYlib-MUQ} interface \cite{kim_hippylib-muq_2023}.
\section{Results}\label{sec:results}

\subsection{Data Generation and Preprocessing}

In problems involving stochastic processes, the data $\bsy$ is typically not available directly as the (noisy) solution of the MFPT or FP equations. We rather work with ensembles of trajectories, which we assume to be realizations of the underlying process. Suppose now that we have such an ensemble available, e.g. from experimental observations or black box simulations. With regard to the MFPT data, we consider the case of a fixed number of initial sites $\bsx_i\in\Omega,\ i=1,2,\ldots,M$. For each of these initial sites, we further assume w.l.o.g. a fixed-size ensemble of trajectories $j=1,2,\ldots,N$ originating from that point. Then for each trajectory $j$ starting at $x_i$, we can extract an exit time $\tau_j^{(i)}$ from the domain $\Omega$. We now employ simple Monte Carlo estimates for the empirical MFPT moments. For the current application, we restrict ourselves to the first and second moment. Furthermore, for every initial site, we split up the trajectories into sub-ensembles of size $N_1 = N_2 = N/2$ (assuming N to be even) to avoid correlation between the estimates. We obtain the unbiased estimates
\begin{equation}
    \hat{\tau}_1^{(i)} = \frac{1}{N_1}\sum_{j=1}^{N_1}\tau_j^{(i)},\quad \hat{\tau}_2^{(i)} = \frac{1}{N_2}\sum_{j=1}^{N_1}\big(\tau_j^{(i)}\big)^2.
\end{equation}
In addition, we model the pointwise standard deviation in the data noise as the standard error of the respective estimators \cite{kroese_handbook_2011},
\begin{equation}
    \hat{\sigma}_\alpha^{(i)} = \frac{S_\alpha^{(i)}}{\sqrt{N_\alpha}},\quad \big(S_\alpha^{(i)}\big)^2 = \frac{1}{N_\alpha-1}\sum_{j=1}^{N_\alpha}\Big( \big(\tau_j^{(i)}\big)^\alpha - \hat{\tau}_\alpha \Big)^2,\quad \alpha=1,2.
\end{equation}

Next, we discuss data preprocessing for inference based on the FP equation. Here, we assume an ensemble of $N$ trajectories with some known initial distribution $p_0$. From this ensemble, we extract snapshots at times $t_i>0,\ i=1,2,\ldots,M$. For each of the snapshots, we compute an estimate for the PDF at that time via Kernel density estimation with a Gaussian kernel \cite{silverman_density_1992,tsybakov_introduction_2009},
\begin{equation}
    \hat{p}_N(x, t_i) = \frac{1}{Nh} \sum_{j=1}^N K\bigg(\frac{X_j^{(i)}-x}{h}\bigg),\quad K(x) = \frac{1}{2\pi}\text{e}^{-\frac{x^2}{2}}.
\end{equation}
We manually choose the bandwidth $h$ such that the bias of the KDE estimator is small compared to its variance, allowing for the assumption of unbiased data. Specifically, we require that
\begin{equation}
    \text{bias}=\calO(h^2) \ll \text{variance}=\calO\Big(\frac{1}{Nh}\Big).
\end{equation}
We can further compute an estimate for the pointwise KDE variance as
\begin{equation}
    \hat{\sigma}_i^2(x) = \frac{1}{Nh}\hat{p}_N(x,t_i)\sigma_K^2,\quad \sigma_K^2 = \int K^2(y)dy.
\end{equation}

\subsection{Single-Scale Process}
\begin{figure}[ht]
    \centering
    \begin{subfigure}{0.35\textwidth}
        \includegraphics[width=\textwidth]{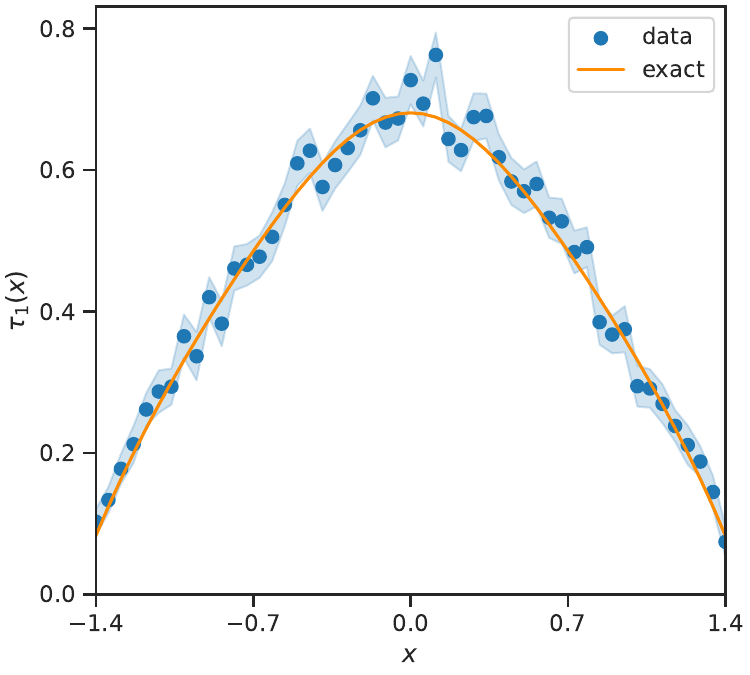}
    \end{subfigure}
    \begin{subfigure}{0.35\textwidth}
        \includegraphics[width=\textwidth]{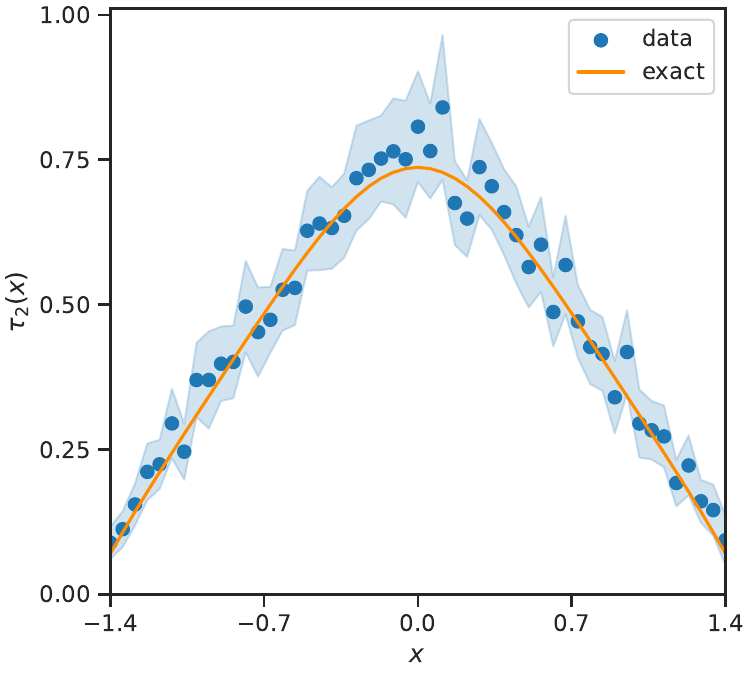}
    \end{subfigure}
    \caption{Data from Monte Carlo estimates of the first and second MFPT moments, together with the FEM solution for the true drift and diffusion functions. Colored intervals indicate pointwise $1.96$ standard deviations from the mean.}
    \label{fig:mfpt_data}
\end{figure}

\begin{figure}[ht]
    \centering
    \begin{subfigure}{0.32\textwidth}
        \includegraphics[width=\textwidth]{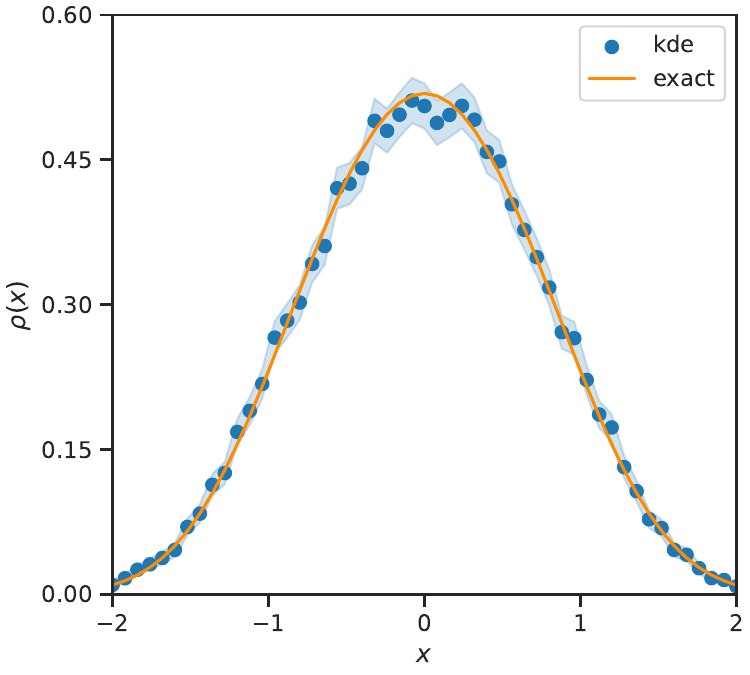}
    \end{subfigure}
    \begin{subfigure}{0.32\textwidth}
        \includegraphics[width=\textwidth]{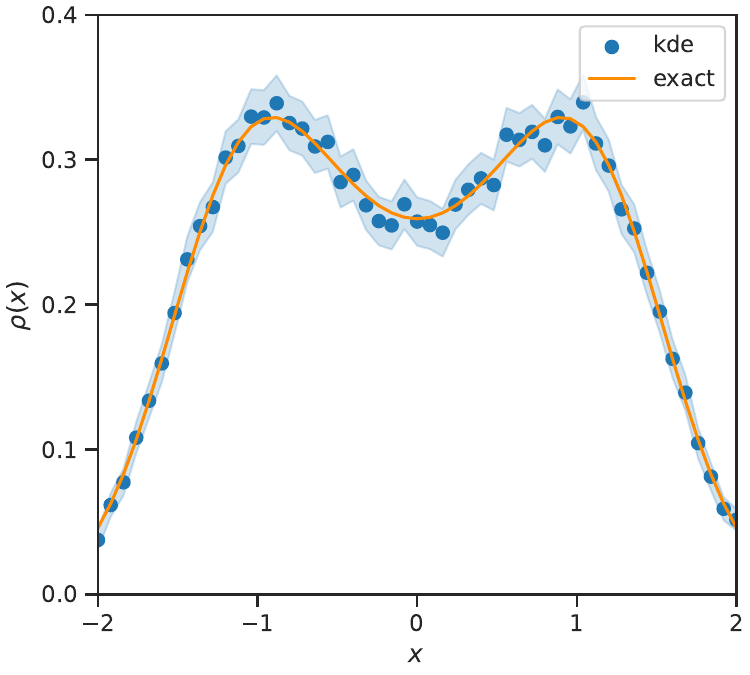}
    \end{subfigure}
    \begin{subfigure}{0.32\textwidth}
        \includegraphics[width=\textwidth]{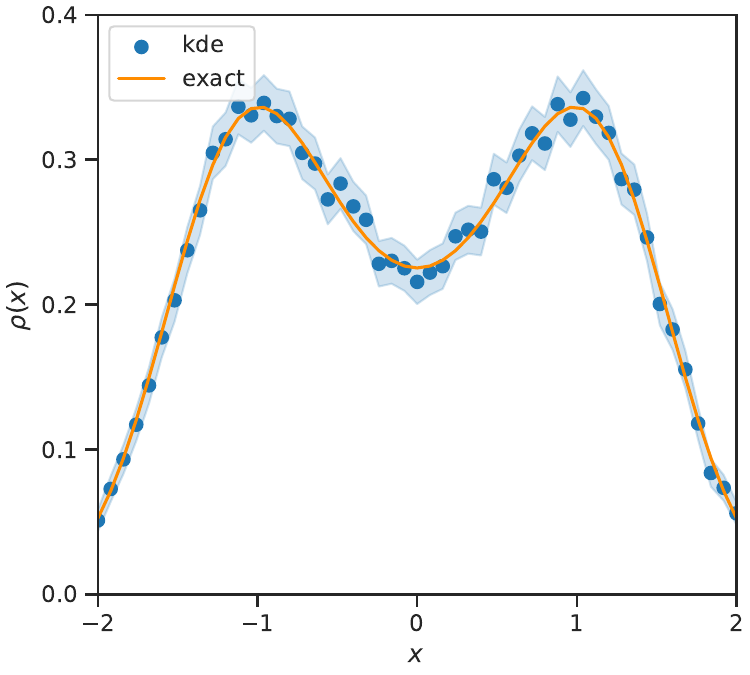}
    \end{subfigure}
    \caption{Data from kernel density estimates of the pdf for $t=\{0.1, 0.5, 0.9\}$, together with the FEM solution for the true drift and diffusion functions. Colored intervals indicate pointwise $1.96$ standard deviations from the estimator.}
    \label{fig:fp_data}
\end{figure}

As a first proof-of-concept study, we consider a one dimensional process with cubic drift and quadratic diffusion term,
\begin{equation}
    b(x) = -2x^3 + 3x,\quad \sigma^2(x) = x^2 + 2.
\end{equation}
To generate MFPT data, we numerically integrate the SDE \eqref{eq:ito_sde} with the Euler--Maruyama (EM) scheme. We compute $10^3$ trajectories at 51 initial sites each, performing a total of 500001 steps with a step size of $10^{-4}$. The obtained data for the first two MFPT moments is depicted in figure \ref{fig:mfpt_data}. Similarly, we generate for the FP data an ensemble of $10^5$ trajectories, with initial states distributed as $\calN(0,\frac{1}{2})$. We perform 1001 EM steps of size $1\mathrm{e}{-4}$. As the final data sample, we extract snapshots in time intervals of $0.1$, of which three are shown in figure \ref{fig:fp_data}.

With the given data sets, we conduct Bayesian inference for the drift and diffusion function of the stochastic process according to the procedures outlined in sections \ref{sec:methods} and \ref{sec:numerics}. Importantly, we do not infer the diffusion function $\sigma(x)>0$ directly, but rather its log square $\log(\sigma^2(x))\in\bbR$. This is to ensure positivity of $\sigma$ and remove ambiguity in the inference through the square root.
As the prior measure, we employ a standard Ornstein--Uhlenbeck (OU) process, i.e. $b(x)=-x$, $\log(\sigma^2(x))=1$.
For both the MFPT and FP models, we construct a Laplace approximation of the posterior, followed by MCMC sampling based on the MALA proposal.

\begin{figure}[ht]
    \centering
    \begin{subfigure}{0.35\textwidth}
        \includegraphics[width=\textwidth]{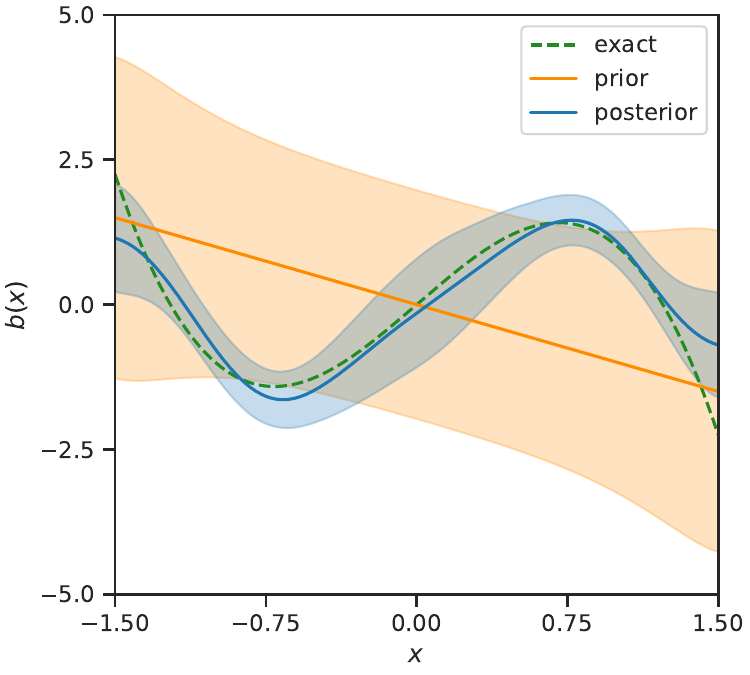}
    \end{subfigure}
    \begin{subfigure}{0.35\textwidth}
        \includegraphics[width=\textwidth]{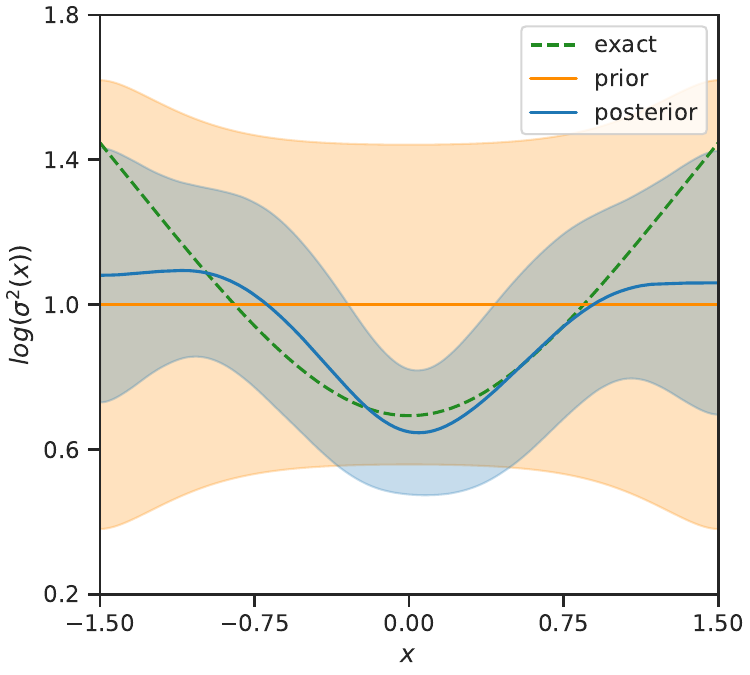}
    \end{subfigure}

    \hspace{0.05cm}
    \begin{subfigure}{0.34\textwidth}
        \includegraphics[width=\textwidth]{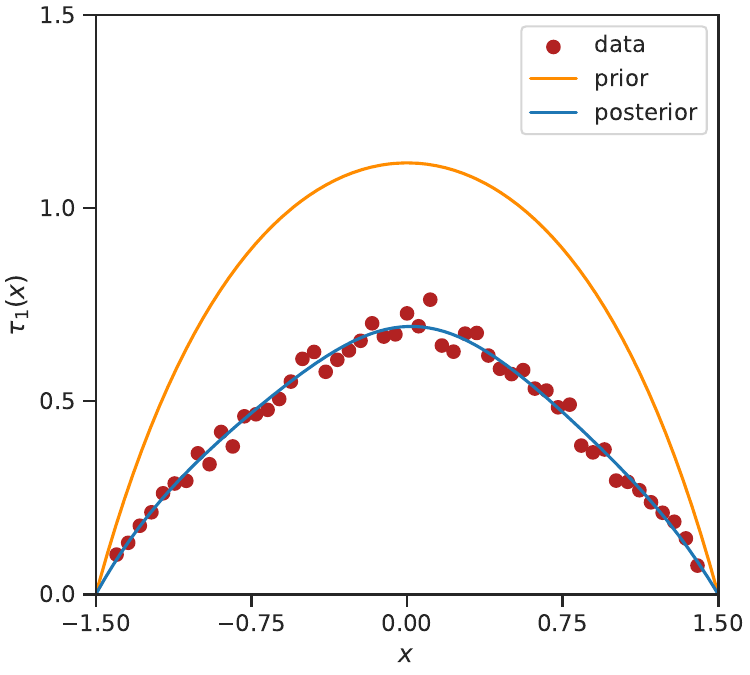}
    \end{subfigure}
    \hspace{0.05cm}
    \begin{subfigure}{0.34\textwidth}
        \includegraphics[width=\textwidth]{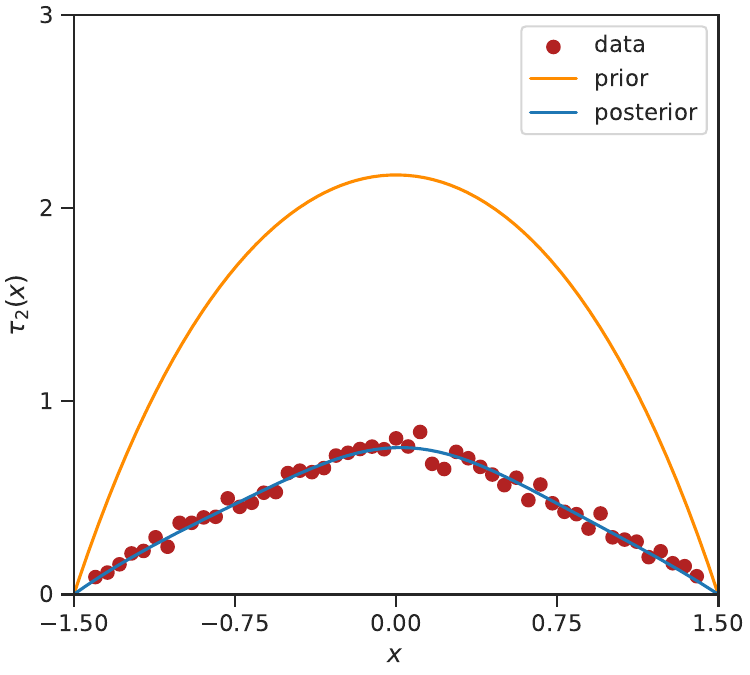}
    \end{subfigure}
    \caption{Top row: Posterior mean for the drift and log squared diffusion function, obtained through inference from MFPT moments data, and compared to the prior and exact solution. Colored intervals indicate $1.96$ standard deviations from the respective mean. Bottom row: Posterior mean predictive for the first two MFPT moments, compared to the prior mean predictive and the utilized data points.}
    \label{fig:mfpt_inference}
\end{figure}

Figure \ref{fig:mfpt_inference} displays the result of the inference from MFPT data, along with the posterior predictive. We observe a good correspondence between the posterior mean and the exact solution for both the drift and the diffusion function. The posterior shows a significant reduction in variance compared to the relatively uninformed OU prior. The estimation of the drift shows less uncertainty, as the diffusion function inference appears to be more sensitive with respect to noise in the data.
Moreover, deviations from the true solution are generally more pronounced towards the domain boundaries. This is to be expected, as the PDE solution is less influenced by the parameter functions close to Dirichlet boundaries. Lastly, we point out  that the posterior predictive shows excellent agreement with the data, indicating that its information content is effectively utilized in the inference.

\begin{figure}[ht]
    \centering
    \begin{subfigure}{0.32\textwidth}
        \includegraphics[width=\textwidth]{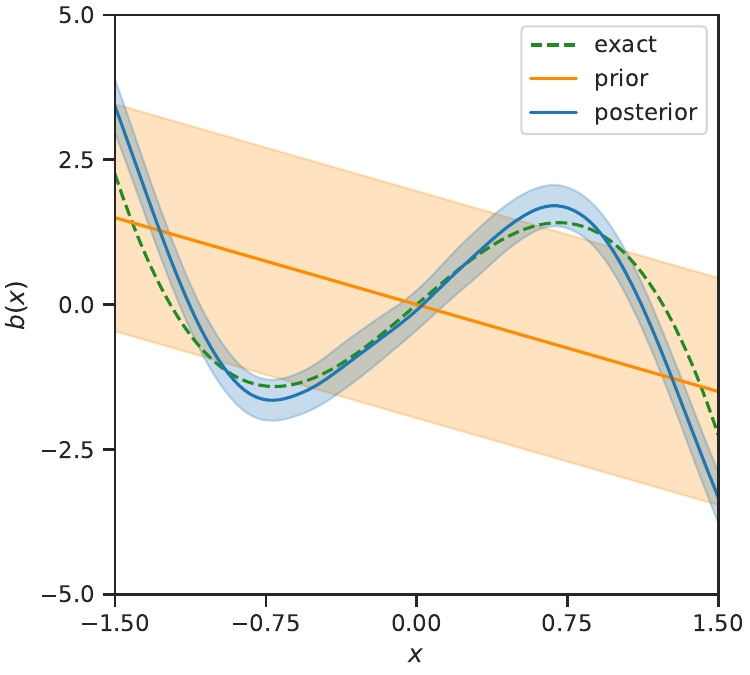}
    \end{subfigure}
    \begin{subfigure}{0.32\textwidth}
        \includegraphics[width=\textwidth]{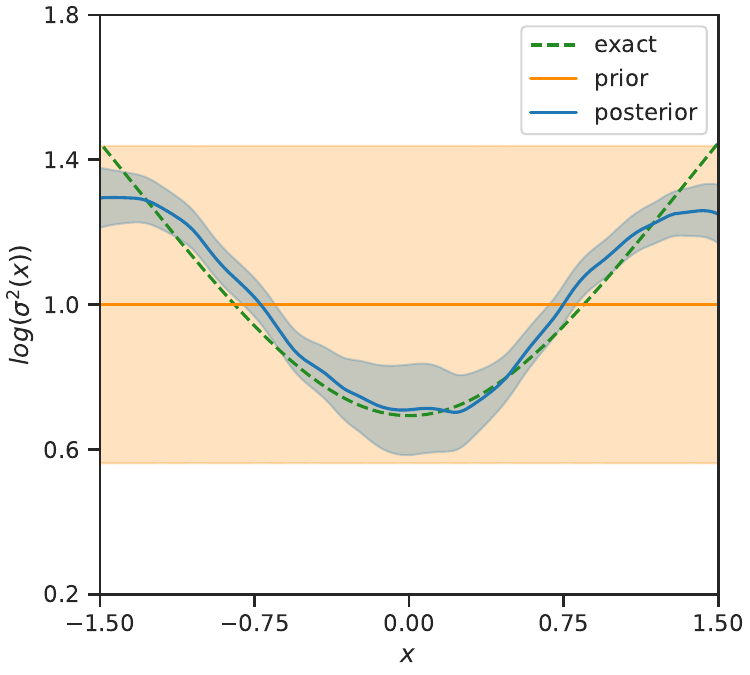}
    \end{subfigure}
    \begin{subfigure}{0.32\textwidth}
        \includegraphics[width=\textwidth]{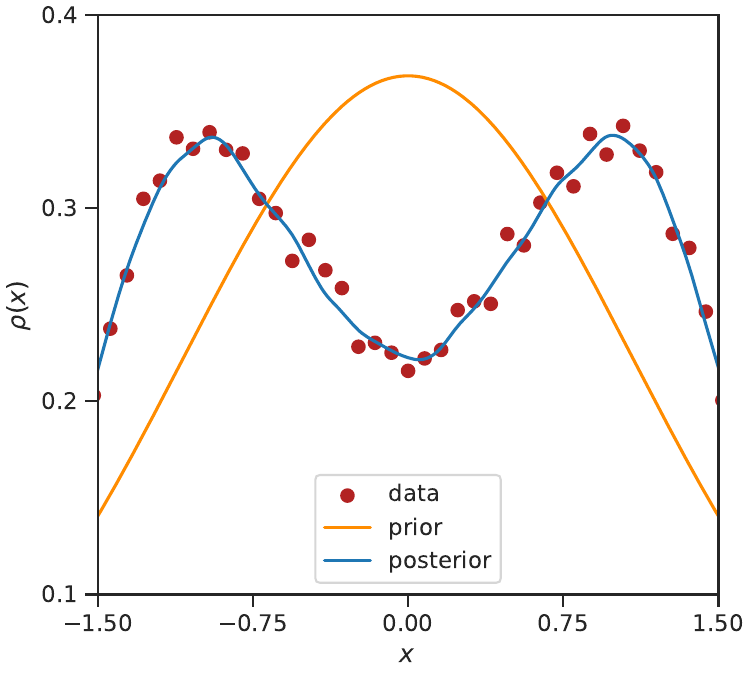}
    \end{subfigure}
    \caption{Posterior mean for the drift (left) and log squared diffusion (middle) function, obtained through inference from FP data, and compared to the prior and exact solution. Colored intervals indicate $1.96$ standard deviations from the respective mean. Posterior mean predictive at $t=0.9$ (right), compared to the prior mean predictive and the utilized data snapshot.}
    \label{fig:fp_inference}
\end{figure}

We observe even better agreement with the true solution for inference based on FP data, as shown in figure \ref{fig:fp_inference}. Again, this is expectable, as we employ multiple snapshots in time, of which each comprises a set of points in space. Variance reduction is also more pronounced, resulting in an even slightly over-confident estimate. Larger deviations towards the domain boundaries occur for the same reason as for the MFPT inference. The posterior predictive aligns perfectly with the utilized data.

\FloatBarrier
\subsection{Multi-Scale Process}

\begin{figure}[ht]
    \centering
    \begin{subfigure}{0.32\textwidth}
        \includegraphics[width=\textwidth]{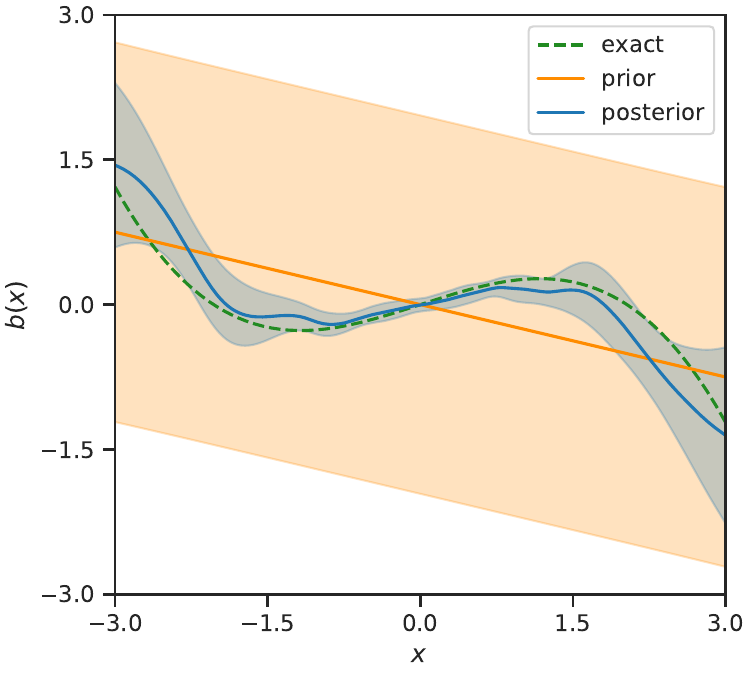}
    \end{subfigure}
    \begin{subfigure}{0.32\textwidth}
        \includegraphics[width=\textwidth]{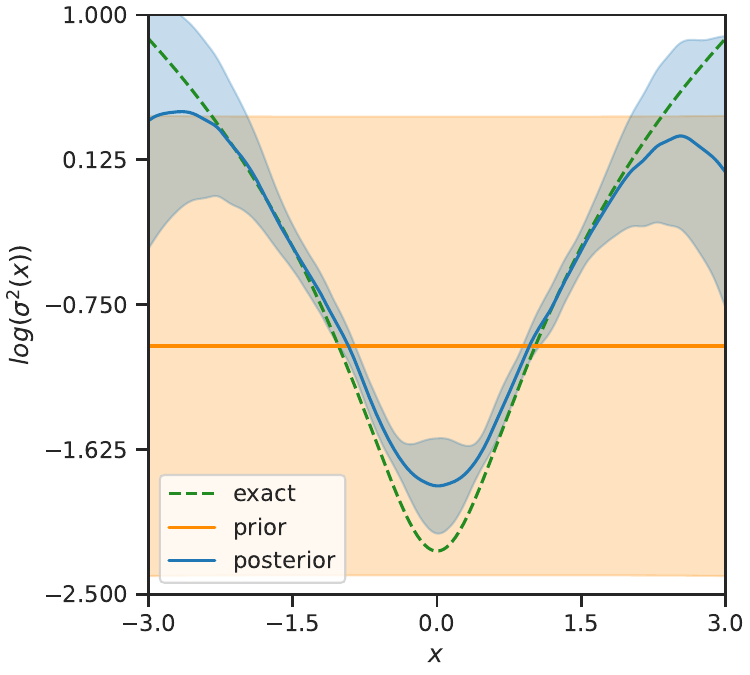}
    \end{subfigure}
    \begin{subfigure}{0.32\textwidth}
        \includegraphics[width=\textwidth]{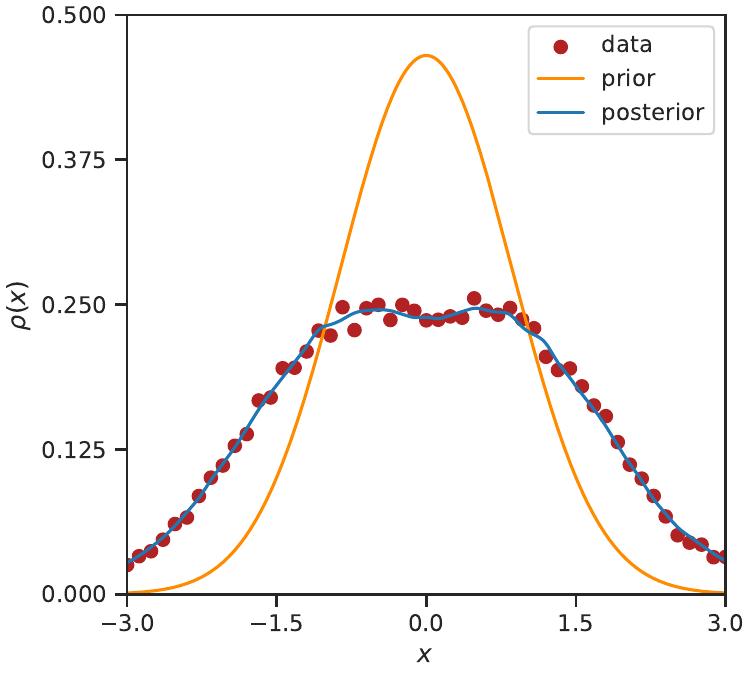}
    \end{subfigure}
    \caption{Posterior mean for the drift (left) and log squared diffusion (middle) function, obtained through inference from multiscale FP data, and compared to the prior and exact solution for the coarse-grain model. Colored intervals indicate $1.96$ standard deviations from the respective mean. Posterior mean predictive at $t=0.9$ (right), compared to the prior mean predictive and the utilized data snapshot.}
    \label{fig:fp_inference_multiscale}
\end{figure}

As a second example, we consider a stochastic process encompassing multiple time scales. We define a system of equations of stochastic variables \cite{krumscheid_semiparametric_2013},
\begin{subequations}
\begin{align}
    dx_t &= \Big( \nu x_t - \frac{1}{2\epsilon}(x_ty_t+y_tz_t) \Big)dt, \\
    dy_t &= \Big( \nu y_t - \frac{3}{\epsilon^2}y_t - \frac{1}{2\epsilon}(2x_tz_t-x_t^2) \Big)dt + \frac{q_1}{\epsilon}dV_t^1, \\
    dz_t &= \Big( \nu z_t -\frac{8}{\epsilon^2}+ \frac{3}{2\epsilon} \Big) + \frac{q_1}{\epsilon}dV_t^2,
\end{align}
\label{eq:multiscale_system}
\end{subequations}
with $q_1=q_2=\nu=1$, and $V_t^1, V_t^2$ being independent standard Brownian motions. The parameter $\epsilon \ll 1$ induces a separation of time scales between $x_t$ (slow) and $y_t,z_t$ (fast). In the limit $\epsilon\to 0$, $x_t$ attains an effective, coarse-grain representation,
\begin{subequations}
\begin{gather}
    dX_t = (AX_t - BX_t^3)dt + \sqrt{\sigma_a+\sigma_bX_t^2} dW_t, \\
    \text{with}\quad A = \nu + \frac{q_1^2}{396} + \frac{q_2^2}{352},\quad B = \frac{1}{12},\quad \sigma_a = \frac{q_1^2q_2^2}{2112}, \quad \sigma_b = \frac{q_1^2}{36}.
\end{gather}
\end{subequations}
We intend to infer the parameter functions of this effective diffusion process $X_t$ from multiscale data of $x_t$. For a preliminary result, we restrict the exposition to the inference approach via the FP equations. Analogously to the single-scale example, we numerically integrate the equation system \eqref{eq:multiscale_system} with the EM scheme, for a parameter value $\epsilon = 0.1$. We generate $10^5$ trajectories, integrating for 50001 steps with a step size of $10^{-3}$. We again utilize a standard OU process for the prior measure, and conduct variational inference via the Laplace approximation. The resulting posterior, as well as the posterior predictive, is displayed in figure \ref{fig:fp_inference_multiscale}. These results are similar to the ones obtained for the single-scale process. We can effectively recover the latent dynamics from the multiscale data, with the posterior predictive neatly aligning with the observations. Interestingly, we can recover the asymptotic dynamics for $\epsilon\to 0$, despite the relatively large value $\epsilon=0.1$ for the generation of the multiscale data. This highlights the importance of the prior measure as a function space regularizer.

\section{Conclusion}\label{sec:conclusion}

In this work, we have presented a method for non-parametric Bayesian inference of the parameter functions of diffusion processes. We have based our forward model on the processes' generators via the Kolmogorov PDEs, which describe the spatio-temporal evolution of observables or probability densities of an underlying process. From the governing PDEs, we have formulated a likelihood and prior measure in a function space setting. Subsequently, we have discussed solution methods for the (consistently) discretized BIP, based on optimization and sampling techniques for large-scale problems.

We have conducted preliminary studies with simulated trajectory ensembles for a single-scale and a multiscale stochastic process. Our results indicate that a fully non-parametric inference of both drift and diffusion functions is feasible. This is also the case for learning an effective coarse-grain representation from multiscale trajectory data. In addition, we point out that the inference procedure appears to be reasonably robust with respect to the noise level of the available data. On the other hand, our approach requires a substantial amount of trajectory data to approximate the required observables. It is therefore more suitable for inference on black box simulators than for experimental data. Other work typically conducts inference from single long-time trajectories by constructing a likelihood from the transition probabilities of a Markov process \cite{croix_nonparametric_2020,crommelin_estimation_2012,papaspiliopoulos_nonparametric_2012,van_der_meulen_reversible_2014,krumscheid_data-driven_2015,dietrich_learning_2023}. However, these approaches appear to be less powerful, in the sense that they are not fully non-parametric, make stronger a-priori assumptions or do not simultaneously infer drift and diffusion.

As a possible future work package, we envision the application of our approach to learning the latent dynamics of many-particle systems. This would require the extension towards non-Markovian processes and PDEs with convolutions. Regarding methodology, we point out that we have followed the simplest workflow possible for creating and processing the necessary trajectory data. More elaborate methods could potentially reduce the required amount of data for accurate inference.

\section*{Acknowledgements}
The authors acknowledge support by the state of Baden-Württemberg through bwHPC. Furthermore, S.K.\ acknowledges funding by the Deutsche Forschungsgemeinschaft (DFG, German Research Foundation) - Project number 442047500 through the Collaborative Research Center "Sparsity and Singular Structures" (SFB 1481).

\newpage
\bibliographystyle{spmpsci}
\bibliography{references}

\end{document}